# Building the National Radio Recordings Database: A Big Data Approach to Documenting Audio Heritage

Emily Goodmann
Dept. of Business,
Communication, and CIS
Clarke University
Dubuque, Iowa, USA
emily.goodmann@clarke.edu

Mark A. Matienzo
Stanford University Libraries
Stanford University
Stanford, California, USA
orcid.org/0000-0003-3270-1306
matienzo@stanford.edu

Shawn VanCour
Dept. of Information Studies
University of California,
Los Angeles
Los Angeles, California, USA
svancour@ucla.edu

William Vanden Dries
Archives of African American
Music and Culture
Indiana University
Bloomington, Indiana, USA
wvandend@indiana.edu

*Abstract*—This paper traces strategies used by the Radio Preservation Task Force of the Library of Congress's National Recording Preservation Board to develop a publicly searchable database documenting extant radio materials held by collecting institutions throughout the country. Having aggregated metadata on 2,500 unique collections to date, the project has encountered a series of logistical challenges that are not only technical in nature but also institutional and social, raising critical issues involving organizational structure, political representation, and the ethics of data access. As the project continues to expand and evolve, lessons from its early development offer valuable reminders of the human judgment, hidden labor, and interpersonal relations required for successful big data work.

## I. A National Radio Recordings Database

In 2014, the National Recording Preservation Board (NRPB) of the United States Library of Congress created a Radio Preservation Task Force (RPTF) to develop a national strategy for documenting and preserving nearly a century's worth of recorded radio materials held by archiving institutions throughout the country. As part of its activities, the RPTF was charged with conducting a comprehensive survey of extant collections and aggregating collections data received from contributing archives into a publicly searchable online database [1]. This paper discusses the development of this national radio recordings database and the various technical, organizational, and political and ethical issues involved in designing and executing this first-of-its kind big data project.

As a big data project, the national radio recordings database has encountered several challenges. The geographically distributed nature of US radio archives and lack of any centralized national repository presented logistical challenges for initial data gathering efforts and has yielded a wide range of participating collecting institutions with varying levels and types of collection metadata. At the level of metadata aggregation, this diversity of source material has created distinct challenges for data management and standardization, while the need to construct a dynamic, living database whose content is continually expanded and updated has presented further complications distinct from more static models of dataset design. Organizationally, the project has demanded strategic recruitment of appropriate expertise, development of and coordination between appropriate divisions within the RPTF, and communication across multiple disciplinary sectors. The project has additionally raised a series of political and ethical issues that offer important reminders of the "human" side of big data. Politically, ingrained in survey methods and dataset design are human decisions impacting whose audio heritage is prioritized within the context of a "national" preservation project, while ethically the team has also encountered several critical privacy issues for collections whose custodians wish to selectively withhold some or all details about their holdings. Throughout the process, the human labor undergirding a big data project has also been repeatedly foregrounded – in questions surrounding resource allocations within the RPTF, in judgments made by the RPTF's survey and database teams, and in ongoing work with the hundreds of librarians, archivists, and collectors whose ability to contribute data to the project has been impacted by a wide range of labor issues and working conditions within their own institutions.

## II. Project Origins

### A. NRPB Mandate and Initial Collections Survey

The Radio Preservation Task Force was created following a year-long collections survey conducted by the NRPB. Charged by the United States Congress under the National Recording Preservation Act of 2000 with "implement[ing] a comprehensive national sound recording preservation program" [2], the NRPB published its official National Recording Preservation Plan in December 2012. Radio content, the Board noted in this document, posed special challenges, since "many libraries and archives have acquired collections of historical radio broadcast recordings," but "there have been few systematic efforts to … document and preserve the entire range of extant broadcasts" [3]. To develop a more complete picture of the scope and state of these collections, in Summer 2013 the Board organized a first-of-its-kind national radio collections survey conducted by a team of one hundred content specialists recruited from research universities across the country. Initial survey work began in Fall 2013 and ran for one year, with results including data from collecting institutions ranging from state and federal archives to university libraries, state broadcasting organizations, individual radio stations, local historical societies and preservation groups, and private collectors, which together provided evidence of more than 640 extant collections. However, with reports from the survey team indicating the presence of additional, undocumented collections and requests for further time from archivists seeking to produce more complete information on their holdings, the initial survey window was extended for a second year, with scheduled completion in Fall 2015. To supervise these research activities and advise the Board on relevant preservation issues, the NRPB created a Radio Preservation Task Force led by broadcast historians Chris Sterling (George Washington University) and Josh Shepperd (Catholic University of America), and charged the group with compiling survey results into "an online inventory of extant American radio archival collections" that would serve as a publicly accessible research database [4]. Survey findings were discussed at a February 2016 conference, and work on the database began shortly thereafter under the direction of RPTF Metadata Director William Vanden Dries (Indiana University). The Metadata Division launched the initial prototype in May 2016 but required several strategic improvements, with co-director Mark A. Matienzo (Stanford University) joining the team to assist in this effort in Fall 2017, and the two bringing a redesigned version online in Spring 2018.

*B. Repercussions for Database Work*

This circumlocuitous project trajectory had several repercussions for resulting database work that warrant consideration from critical data studies perspectives. First, as critical infrastructure scholars note, while there is a tendency for infrastructure to be laid on prior infrastructure, new layers are designed for functions and uses that often diverge from those of earlier layers, creating interoperability problems whose solution demands "moving between the separate registers of technical and social action" [5]. Initial survey design did not anticipate the imagined end goal of an online database, demanding several post-hoc changes in operating procedures that had to be developed and deployed in situ. Issues addressed by the metadata team have included technical questions surrounding standardization of data fields and development of controlled vocabulary for information initially collected using more open-ended or semi-structured research methods, as well as larger privacy concerns surrounding public visibility of collections data initially provided in confidence.

Second, while original collections surveys were conducted by faculty researchers selected largely on the basis of content expertise, the newly created Task Force was designed as a cross-sector project with a mandate to encourage "collaboration between faculty researchers and archivists toward the preservation of radio history." [4] This approach has demanded often difficult processes of disciplinary translation between information scholars, scholars with humanities-based training, and archiving professionals. However, it has also revealed a common, interdisciplinary commitment to inclusion of collections documenting experiences of traditionally marginalized social groups, with shared ideals shaped by critical cultural approaches in broadcast historiography [6] and critical archival studies in the archival sciences [7]. While this activist archiving ethos has facilitated inclusion of many smaller, nontraditional collecting institutions and private collectors whose radio holdings would otherwise be omitted, the collections data maintained by these institutions and levels of archival expertise within them has varied widely, compounding problems of data quality and standardization.

Finally, while databases have traditionally been treated as inert objects decontextualized from their moments of production, critical data studies reminds us that these technical artifacts remain products of often invisible human labor and judgment that are in turn shaped by institutionally specific mandates and organizational procedures [8]. The shift from an initial year-long collections survey to an ongoing online database project necessitated not only technical strategies for a dynamic database architecture that could accommodate regular updates, but also new operational procedures governing the human agents charged with collecting and processing this data. Organizationally, the RPTF established a special unit to conduct outreach work with existing affiliate archives and solicit collections information from additional archives not currently within the RPTF network. First created in 2015 and now run by Network Director Emily Goodmann (Clarke University), this Network Division has coordinated closely with the Metadata Division to develop and refine a series of operational procedures to supply the Metadata team with needed data on new collections. Successful deployment of the National Radio Recordings Database has in this sense thus demanded that flexible application and dataset design have a corresponding flexible institutional design within the RPTF, allowing it to remain adaptable to shifting personnel and evolving organizational needs.

III. DATA COLLECTION AND DATABASE DEVELOPMENT

*A. Collection and Aggregation of Survey Data*

For its original collection survey conducted in 2013-2014, the survey team took a regional approach, with three research teams assembled into Eastern, Western, and Midwestern Divisions led by regional directors reporting to national research director Shepperd. Within their regions, these directors allocated territories for their research teams, and instructed these teams to seek out potential repositories of radio recordings and document information on their collections. While each region was charged with collecting the same basic information, the exact survey methods and questions remained the responsibility of the respective research teams and often adapted within a single region to suit the needs of individual collecting institutions. Some respondents filled out online Google Forms that fed data into spreadsheets, others returned paper sheets, while still others participated in phone interviews, relayed relevant collections information via email communications, or sent in entire finding aids that were then interpreted and logged by researchers and passed up to division heads, who compiled this data for the national research director for assembly into the comprehensive report delivered to the NRPB in December 2014.

With the formation of the RPTF and issuance of the NRPB's database mandate, newly appointed Metadata Director Vanden Dries began the work of combining all of the information used for the NRPB report and transitioning it into an online database, while the RPTF's new Network Division continued soliciting collections data from additional collecting institutions. Existing data was compiled into an Excel workbook, digitizing paper-based survey responses in the process. To transform the data into an online resource, the information was reviewed to generate a set of database fields based on the questions asked in the survey. The conclusion was to use the fields found in Encoded Archival Description (EAD) for collection-level description, with several additional fields to store and display survey data that did not fit easily into this existing field set. The data compilation process revealed many voids in the descriptive data due to the inconsistencies in the initial survey questions. A second round of surveying was initiated with the organizations and individuals that returned collection information to fill as many of these information gaps as possible, using researchers under the direct supervision of the Metadata Director. During this re-surveying process, multiple respondents also sent new collection information not included in their first response.

### B. Development of the Database Application

After reviewing various search and discovery platforms, the RPTF decided to move forward with an application built using the open source Blacklight framework [9], which uses Ruby on Rails, MySQL, and Apache Solr among other software packages to support searching and presentation of metadata. Originally developed by the University of Virginia to support discovery of bibliographic resources cataloged using MARC21 [10], Blacklight supports the searching and presentation of any dataset indexable into Solr. The selection of Blacklight was a strategic choice for the creation of the database given its widespread use by information organizations for search and discovery. A second reason is the ongoing development of ArcLight [11], an extension to Blacklight offering options for more fine-grained, hierarchically ordered archival description below the collection level. The RPTF concluded that if ArcLight development continued in the same direction, the public interface would be updated to the ArcLight interface in the future. The Association for Recorded Sound Collections (ARSC), an organization closely tied to radio preservation and influential in the creation of the RPTF, was approached for assistance with web hosting and agreed to fund the costs necessary to deploy the site by expanding the hosting package it used for its own website to include a virtual private server (VPS) dedicated to the radio collections database. With web hosting space secured, the Metadata Director began development of the Blacklight platform, with additional rounds of data cleanup completed as needed and the prototype brought online for the first time in May 2016. Documentation regarding the development of the initial version of the database was created and maintained by the Metadata Director [12].

### C. Coordinating with RPTF Expansion and Outreach Activities

To streamline coordination with the RPTF's newly established Network Division, the Metadata Division created a public-facing Google Form for entering information on new collections, which included predefined data fields and a consistent vocabulary control scheme designed to facilitate easy ingestion of additional collection information into the expanding online database. The Network Team pointed collection holders to this form, which received and temporarily held the data until it could be reviewed by the Metadata Team, augmented if necessary, with supplemental information, and added to the Solr index feeding the search and discovery interface. This effectively bridged the two divisions of the organization, while relegating data solicitation activities by the Network Division to an outreach role and allowing direct control of resulting metadata by members of the Metadata team. To further grow its existing list of extant collections, the Metadata Division developed a system for incorporating collection data already aggregated in print publications, using as a well-known reference work in radio history [13] as a test case. This reference work was scanned and had machine-readable text produced using optical character recognition (OCR). This proved a time-consuming process that required extensive cleaning by researchers to correct OCR errors, extract relevant data, then manually format it to align with the Solr index fields; however, the approach added several hundred collection-level records beyond those represented in the original survey, pushing the total to over 1,000, with continuing outreach efforts by the Network Team in turn adding several hundred further collections to the original list within the database's first year of operation.

This rapid growth, while a salutary development, also created several problems for the original database design. Usability and stability updates were necessary, but the available human labor and technical knowledge of the existing Metadata Team created impediments to designing and implementing the needed changes. To resolve these issues, the RPTF directorate recruited Matienzo as second Metadata Director to bring more in-depth knowledge of metadata aggregation, as well as ArcLight and Blacklight, while other team members focused on completing data aggregation and cleanup of existing records for incorporation into the database.

## IV. IMPROVEMENTS AND GROWING PAINS

Following the recruitment of the second Metadata Director, the Network and Metadata Divisions held a special session at a second RPTF conference convened in November 2017, where they developed a strategic plan [14] for improving the database's interface and stability, as well as expanding the number of collections represented within it in response to pressure from the NRPB to advance the project as rapidly as possible. This plan was further developed in the months immediately following the conference and has been systematically implemented throughout 2018 and 2019. While interface and functionality issues have proven straightforward technical problems, continued expansion of the database has introduced issues that cannot be solved through technical fixes alone but also raised significant cultural and ethical questions.

### A. Technical Improvements

Work to improve the database's interface and functionality included updating Blacklight, Solr, and other dependencies. The Metadata Division has streamlined the ingest process to make its work easier, particularly in terms of both adding and deleting records. While adoption and incorporation of ArcLight was expected as a potential work item for the Metadata Division, the database's focus on collection-level description meant that ArcLight was a poor match, since one of ArcLight's core features is to present hierarchical archival description [15]. While early plans for the database included the possibility for presenting full hierarchical finding aids, the significant variation in the collections information received and compiled by the Network and Metadata Divisions led the Metadata Directors to deemphasize the inclusion of ArcLight. Nonetheless, additional efforts are underway to improve the database's functionality. One notable area of interest is providing better curatorial tools to group, contextualize, and present additional information around potential collections. In its simplest form, the RPTF expects this would take the form of hosting educational guides and user-created finding aids or providing links to other external resources when available. An additional possibility is looking at expanding the database to incorporate the Spotlight framework [16], a Blacklight extension originally developed by Stanford University Libraries to create customized exhibits that would provide a native administrative interface within the database to maintain or alter the collections information, allowing for smoother dynamic updates to existing entries as collections grow or additional processing work is completed by the contributing archives. The Metadata Division is also pursuing continuing efforts to provide expanded search functionality, including refining search facets used as entry points or to refine queries in the database, and exposing structured data for items in the database and search results in JSON through an application programming interface for prospective use by external clients for large-scale data harvesting.

### B. Privacy Issues

To expand the number of collections represented within the database, the Network Division initially focused on private collectors and larger repositories with high volumes of well documented radio materials; however, both sources of collection data presented significant challenges that moved beyond the realm of data science to include ethical and cultural concerns. Among the core concerns noted by information scholars surrounding privacy are the public dissemination of information that individuals wish to remain private, and secondary use of information for purposes beyond what was initially intended [17]. Both sets of concerns have historically loomed large for private media collectors, who for reasons of trade advantage and fears of potential legal prosecution or theft of their materials often prefer to keep many details about their collections hidden from public view [18]. The risk of exposing this private information on a publicly accessible database accordingly made many collectors initially unwilling to participate in the RPTF's project, requiring a combination of technical and social interventions.

To address these privacy concerns, the Metadata Team reconfigured the database ingest process to create filtered subsets that enabled contributors to select between three different sharing settings when supplying information about their collections: (1) all information fully publicly available; (2) share only the collection's owner, its title, and its description publicly; or (3) to restrict it entirely. While the value of "dark" archives is a matter of ongoing debate [19], the RPTF deemed the benefits of documenting these collections while collectors were still alive and willing to cooperate to outweigh the drawbacks of restricted public access. To overcome potential mistrust by private collectors, the RPTF also created a dedicated Program Transcriptions Team staffed by individuals with direct ties to the collecting community, who performed outreach work to convince collectors to contribute information about their collections then referred them to members of the Network Division who entered this information into the relevant sections of the Google Form. These privacy options and trust-building measures were successful, with the RPTF recruiting information from over forty private collectors whose holdings represent a combined total of several hundred thousand unique radio recordings. Planned incorporation of the Spotlight framework in future updates to the database would further enhance its privacy features by allowing even finer item-level designation of specific data fields that contributors wish to either expose or shield from public view, and it is hoped that positive word of mouth among existing collectors who have contributed their information will facilitate further contributions from others within the collecting community.

### C. Strategic Partnerships

Strategic partnerships with larger repositories have presented possibilities for automated harvesting of larger datasets maintained by those institutions, which would offer a swift and expedient way to further grow the RPTF database, following models used by aggregators such as the Digital Public Library of America [20]. Given feedback from NRPB and ARSC, and the continued importance of relationships with both groups, the Metadata Division renamed the database under the advice of fellow directors with a more general name in Spring 2019: the RPTF/ARSC Sound Collections Database. The Metadata Division has also pursued conversations about potential partnerships with Wikidata that would allow for better management of structured data about collection-holding entities. Other opportunities for partnership originate in interest from digital humanities and media history researchers interested in connecting their projects or platforms with the collections information available in the database. These partnerships present an opportunity to provide additional depth to the high-level descriptions that the database provides and may require bi-directional integration of additional computational access methods for the collections information, such as regularly provided dataset snapshots, or an application programming interface.

However, cultivating these partnerships demands not only technical solutions but also extensive human labor, including a common commitment to shared goals commensurate with the RPTF's own institutional mandate and mission statement. While these further lines of development remain largely speculative to date, in the interim the continuing efforts of the Network Division have grown the number of collections within the database to a current total of over 2,500, with an expectation to clear 3,000 in 2020, while the design updates and expanded search functionality implemented since 2018 have enabled a relatively smooth transition from the initial beta version to a fully functioning public research database.

## V. Human Networks

### A. Role of the Network Division

Within the United States, radio archives are geographically dispersed, with no centralized national repository [21], and have varying levels of collection information, competing metadata standards, and divergent levels of data literacy among archival caretakers. The RPTF's Metadata Team has worked to standardize information received from these institutions through its evolving survey form and data cleaning efforts, yet a human interface remains vital to continued expansion of and updates to extant collection data. The RPTF's Network Division plays a vital role in both processes: helping to locate relevant repositories and making the RPTF's mission and methods legible to these prospective data contributors, as well as ensuring collection data is logged in a manner optimized for use by its sister Metadata Division.

### B. Current Data Collection Procedures

The public-facing Google Form used by the RPTF's Metadata Team is the primary survey format used by outreach teams to collect data from individuals and institutions. Problems with non-standardized data in the initial survey process led to the introduction of more closed-ended data fields to enable more consistent vocabulary control. However, the need to balance data precision with user-friendly design has resulted in the persistence of numerous open-ended data fields that allow for a more comprehensive description of collection-level information. To ensure consistent quality of resulting metadata, Network Team members who assist archivists with form completion receive training in relevant data points, and a field definition form that elaborates on desired information for each field is provided to archivists who are pursuing solo entries. Information currently captured includes data regarding ownership, accessibility, archive creators, collection date span, content type, format, physical format of recordings, languages, genres, collection conditions, and collection finding aids. These straightforward data points are supplemented by questions requesting more open-ended inventory descriptions, descriptions of available supporting documentation, historical relevance of the collection, archive access and usage statements, and a free-text field for additional notes not covered in other fields. However, user-friendly design has not obviated the need for human contact, and the Network Team has consistently found that direct communication with archivists concerning the goals and uses of the survey prior to their completion of it produces data that is both more comprehensive and of better quality.

### C. Obstacles to Successful Form Completion

At the same time, managing initial contact and securing successful follow-through on form completion have been persistent challenges. For example, one Network outreach team identified a group of noncommercial station archives as high value targets for database expansion, sending initial outreach emails to over 375 of the 870 archivists charged with managing these collections then scheduling follow-up telephone meetings with those who responded to secure their commitment and explain the survey process. However, heavy workloads and lack of internal support at these archivists' home institutions have resulted in a low success rate to date, with less than half of those who expressed a willingness to complete the form following up with the requested survey data. This offers a useful reminder that the organizational challenges of a big data project such as the RPTF's extend beyond the development of viable procedures for internal coordination among its own divisions to include challenges involving external organizational cultures, workloads, and divisions of labor at the thousands of individual archives that furnish the component collections data.

While survey completion has been challenged by managing hundreds of human contacts within the RPTF's ever-expanding network of contributing archives, it has also been impeded by the survey team's ability to identify and make initial contact with smaller, less visible archives whose holdings remain underrepresented in the database project. The RPTF acknowledges that many of the most valuable collections, from an activist archiving perspective, may be held by smaller community broadcasters and other community-based archives whose materials are more likely to document radio content and listening communities marginalized in traditional histories, and whose limited resources may place those materials at greater risk of deterioration or loss [22]. Those same limited resources can also make successful documentation of these collections particularly challenging, creating a cycle of invisibility that impedes effective preservation and use of these materials. To that end, we believe that the RPTF has much to learn from the growing body of work on community-based archiving [23].

The adoption of a constituent relationship management (CRM) technology to support and manage communication with archivists has become a pressing need. A CRM tool would allow for centralizing and streamlining ongoing communication with archivists, as well as the ability to set reminders for follow-up communication regardless of who is operating or accessing the technology. It would also enforce an ongoing uniformity of survey completion support that will directly impact and increase the quantity of collection-level data within the database. At the same time, however, just as data challenges of the survey collection process have deeper social and economic causes, there are no quick or easy technical fixes to these problems of communication with and follow-through by contributing archives. If "bigger data are not always better data" [24], "better" data might be conceived as data that is not only accurate and consistent but also data yielded through socially just conditions within the preservation institutions and projects responsible for generating it.

## VI. Closing Considerations

### A. Diversifying the Historical Record

As Geoffrey Bowker reminds us, "raw data is both an oxymoron and a bad idea; to the contrary, data should be cooked with care" [25]. The degree to which the RPTF's survey and design methodologies impact the types and visibility of collections within the database is a pressing concern. Beyond a desire for comprehensive record-keeping, the RPTF recognizes its database work as also cultural memory work, with decisions surrounding which types of collections and collecting institutions to privilege or deemphasize determining whose histories and cultural experiences are granted legitimacy and whose are marginalized or suppressed. The survey and design teams have the ambitious goal of ensuring that no extant

American archival radio data is prevented from inclusion in the national database, aiming for a diverse roster of participants that will include institutions and collections that give voice to historically underrepresented communities alongside better-known collections and institutions that speak to more historically dominant sets of identities and cultural experiences [26].

This said, some collections have been easier for survey teams to prioritize than others. Survey and outreach teams began with highly visible noncommercial networks, colleges and universities, and community radio precisely because of their accessibility and, in many cases, public-facing archival holdings. These collections were the so-called "low hanging fruit"—the initial target of any project of this enormity— and were contacted by outreach teams to quickly build an initial database and demonstrate project viability. Moving forward, it will be crucial for survey teams to design and support ongoing outreach to additional private collectors and nontraditional archives with historically underrepresented collections content. The make-up of the task force includes radiophiles of all stripes (collectors, academics, hobbyists, public broadcasters, corporate stakeholders, archivists, historians). Survey teams must continue to communicate with these stakeholders and complete additional research and outreach to locate archives that have not yet been identified in order to prioritize them in their surveying activities.

*B. Human Resource and Technical Issues*

As a service-based organization run on a volunteer basis, the RPTF also faces significant human resource challenges. A revolving door of professional and student research volunteers has been crucial to the success of the survey collection and database creation, but this lack of continuity has also impacted the ability to effectively manage ongoing relationships with archivists and ensure consistency in the data supporting the database. A CRM implementation would help support the quality and quantity of data being indexed in the database about individual and collectors, as well as build better rapport among the project's stakeholders. However, the RPTF has also found it essential to maintain a balance that allows individuals and institutions that hold radio collections to enter information about those collections in a manner that makes the most sense to them.

As this collection information is not static, there is also a need for collection holders to continually add to and modify existing listings. At a technical level, this precludes any definitive or final version of the database and has precipitated ongoing debates concerning the best means of ensuring a dynamic and scalable architecture for the project that can accommodate these ongoing updates and continued expansion. A second lingering technical question concerns preferred means of collecting data from collections holders. Through its revision of the data collection, entry, and transformation processes, the RPTF has focused on adding consistency through both improved survey instruments and stronger personal relationships. To inform both data modeling and discovery concerns, the question remains about who the audiences and stakeholders for the database are, and whether current survey methods will serve the needs of these groups. A clearer picture of its present and desired users would enable the RPTF to take clearer action in improving its data collection, aggregation, and transformation workflows, as well as the overall functionality of the database.

*C. The Humanity of Big Data*

Database projects, once reified, often generate public mythologies that suggest their technical systems are "neutral;" that users can execute queries on their aggregate data to generate unbiased results from a complete set of information [27]. The ongoing challenges related to the RPTF's big data project—survey completion of known archives, locating and facilitating inclusion of underrepresented archives, ethics of data access, building trust with key constituencies, ensuring human resource continuity and successful cooperation within its own organization—remind us that data are far from neutral and, first and foremost, a human artifact. The day-to-day operations of task force members and constituents reveal the complexity and volume of human labor and relationships required to both create data and support its existence across decades. Only later do data become the matter of machines. Ongoing, communicative relationships with archivists, collectors, and data custodians both within and beyond the task force itself will be crucial to continuing to grow an inclusive and comprehensive database that ensures the historical records documented within it remain discoverable to, accessible by, and useful for future generations.

REFERENCES

[1] Radio Preservation Task Force, Sound Collections Database, https://database.radiopreservation.org/ (accessed Oct. 6, 2019).

[2] National Recording Preservation Act of 2000, Public Law No. 106-474. 2000.

[3] "The Library of Congress National Recording Preservation Plan." Washington: Council on Library and Information Resources and the Library of Congress, December 2012.

[4] "Radio Preservation Task Force." Library of Congress. https://www.loc.gov/programs/national-recording-preservation-plan/about-this-program/radio-preservation-task-force/ (accessed Oct. 6, 2019).

[5] S. J. Jackson, P. N. Edwards, G. C. Bowker, C. P. Knobel, "Understanding infrastructure: history, heuristics, and cyberinfrastructure policy," First Monday, vol. 12, no. 6, June 4, 2007, https://journals.uic.edu/ojs/index.php/fm/article/view/1904/1786 (accessed Oct. 6, 2019).

[6] C. Anderson and M. Curtin, "Writing cultural history: the challenge of radio and television," in N. Brügger & S. Kolstrup (eds.), Media History: Theory, Methods, and Analysis,. Aarhus, Denmark: Aarhus University Press, pp. 15-30, 2001.

[7] M. Caswell, R. Punzalan, and T-Kay Sangwand, "Critical archival studies: an introduction," Journal of Critical Library and Information Studies, vol. 1, no. 2, pp. 1-8, 2017.

[8] A. Iliadis and F. Russo, "Critical data studies: an introduction," Big Data and Society, October 2016, pp. 1-7.

[9] "Blacklight: A multi-institutional open-source collaboration building a better discovery platform framework." http://projectblacklight.org/ (accessed Oct. 6, 2019).

[10] E. Sadler, "Project Blacklight: a next generation library catalog at a first generation university," Library Hi Tech vol. 27, no. 1, pp. 57-67, 2009.

[11] M. A. Matienzo, "ArcLight: illuminating discovery to delivery for archives and special collections." Presented at Coalition for Networked Inf. Membership Meeting, Washington, DC, USA, Dec. 11-12, 2017.

[12] W. Vanden Dries, "Blacklight Search/Ruby on Rails App Bluehost Installation." http://rubyonrailsbluehostinstall.blogspot.com/ (accessed Oct. 6, 2019).

[13] S. Siegel and D. S. Siegel, A Resource Guide to the Golden Age of Radio: Special Collections, Bibliography and the Internet. Yorktown Heights, New York: Book Hunter Press, 2006.

[14] "RPTF/ARSC Collections Database 2018 Workplan," unpublished.

[15] "ArcLight," DuraSpace Wiki, https://wiki.duraspace.org/x/94o2BQ (accessed Oct. 6, 2019).

[16] G. Geisler, C. Beer, S. Snydman, J. Keck, and J. Coyne, "Spotlight: a Blacklight plugin for showcasing digital collections." Presented at Open Repositories 2014, Helsinki, Finland, June 9-13, 2014. http://urn.fi/URN:NBN:fi-fe2014070432254 (accessed November 8, 2019).

[17] D. J. Solove, Understanding Privacy. Cambridge: Harvard University Press, 2008.

[18] D. Bartok and J. Joseph. A Thousand Cuts: The Bizarre Underground World of Collectors and Dealers Who Saved the Movies. Jackson, MS: University Press of Mississippi, 2016.

[19] J. R. Baron and N. Payne, "Dark archives and edemocracy: strategies for overcoming access barriers to the public record archives of the future," 2017 Conf. for E-Democracy and Open Government, May 17-19, 2017, doi: 10.1109/CeDEM.2017.27.

[20] M. A. Matienzo, "The Digital Public Library of America's application programming interface and metadata ingestion process," in Putting Descriptive Standards to Work, K. Kiesling and C. Prom, Eds. (Trends in Archival Practice Series). Chicago: Society of American Archivists, 2017.

[21] M. Hilmes and S. VanCour, "Network nation: writing broadcasting history as cultural history," in M. Hilmes (ed.), NBC: America's Network, Berkeley: University of California Press, 2007.

[22] S. VanCour, "Locating the radio archive: new histories, new challenges," Journal of Radio and Audio Media, vol. 23, no. 2, p. 395-403, 2016.

[23] M. Caswell, M. Cifor, and M. H. Ramirez, "'To suddenly discover yourself existing': uncovering the impact of community archives," American Archivist vol. 79, no. 1, pp. 56-81, 2016.

[24] d. boyd and K. Crawford, "Critical questions for big data: provocations for a cultural, technological, and scholarly phenomenon," Information, Communication & Society, vol. 15, 662-669, 2012.

[25] G. Bowker, Memory Practices in the Sciences. Cambridge: MIT Press, 2005, p. 184.

[26] A. Flinn and B. Alexander, "'Humanizing an inevitability political craft': introduction to the Special Issue on Archiving Activism and Activist Archiving," Archival Science, vol. 15, no. 4, pp. 329-335, 2015.

[27] d. boyd, "Undoing the neutrality of big data," Florida Law Review, vol. 67, p. 227, 2016.